\documentclass[twocolumn,english,superscriptaddress,prl,amsmath,amssymb,floatfix,longbibliography]{revtex4-2}
\usepackage[T1]{fontenc}
\usepackage[utf8]{inputenc}
\usepackage{color}
\usepackage{babel}
\usepackage{amsmath}
\usepackage{graphicx}
\usepackage{esint}
\usepackage[unicode=true,pdfusetitle,
 bookmarks=true,bookmarksnumbered=false,bookmarksopen=false,
 breaklinks=false,pdfborder={0 0 0},pdfborderstyle={},backref=false,colorlinks=true]
 {hyperref}

\makeatletter

\usepackage{dcolumn}
\usepackage{bm}
\usepackage{color}
\usepackage{soul}
\usepackage{babel}
\usepackage{amsfonts}
\usepackage{slashed}
\usepackage{enumerate}

\usepackage{babel}
\usepackage{inputenc}

\makeatother

\begin{document}
\global\long\def\sgn{\mathrm{sgn}}%
\global\long\def\ket#1{\left|#1\right\rangle }%
\global\long\def\bra#1{\left\langle #1\right|}%
\global\long\def\sp#1#2{\langle#1|#2\rangle}%
\global\long\def\abs#1{\left|#1\right|}%
\global\long\def\avg#1{\langle#1\rangle}%

\title{Weak-measurement-induced asymmetric dephasing: manifestation of intrinsic
measurement chirality}
\author{Kyrylo Snizhko}
\affiliation{Department of Condensed Matter Physics, Weizmann Institute of Science,
Rehovot, 76100 Israel}
\affiliation{Institute for Quantum Materials and Technologies, Karlsruhe Institute
of Technology, 76021 Karlsruhe, Germany}
\author{Parveen Kumar}
\affiliation{Department of Condensed Matter Physics, Weizmann Institute of Science,
Rehovot, 76100 Israel}
\author{Nihal Rao}
\affiliation{Department of Condensed Matter Physics, Weizmann Institute of Science,
Rehovot, 76100 Israel}
\affiliation{Present affiliation: Arnold Sommerfeld Center for Theoretical Physics,
University of Munich, Theresienstr. 37, 80333 M\"unchen, Germany}
\affiliation{Present affiliation: Munich Center for Quantum Science and Technology
(MCQST), Schellingstr. 4, 80799 M\"unchen, Germany}
\author{Yuval Gefen}
\affiliation{Department of Condensed Matter Physics, Weizmann Institute of Science,
Rehovot, 76100 Israel}
\begin{abstract}
Geometrical dephasing is distinct from dynamical dephasing in that
it depends on the trajectory traversed, hence it reverses its sign
upon flipping the direction in which the path is traced. Here we study
sequences of generalized (weak) measurements that steer a system in
a closed trajectory. The readout process is marked by fluctuations,
giving rise to dephasing. Rather than classifying the latter as ``dynamical''
and ``geometrical'', we identify a contribution which is invariant
under reversing the sequence ordering and, in analogy with geometrical
dephasing, one which flips its sign upon the reversal of the winding
direction, possibly resulting in partial suppression of dephasing
(i.e., ``coherency enhancement''). This dephasing asymmetry (under
winding reversal) is a manifestation of intrinsic chirality, which
weak measurements can (and generically do) possess. Furthermore, the
dephasing diverges at certain protocol parameters, marking topological
transitions in the measurement-induced phase factor.

\emph{}
\end{abstract}
\maketitle

Dephasing is a ubiquitous feature of open quantum systems \citep{Suter2016,Streltsov2017}.
Undermining coherency, it facilitates the crossover to classical behavior,
and comprises a fundamental facet of the dynamics of mesoscopic systems
\citep{Imry1997,Shorter1991,Shimshoni1993,Marquardt2004,Polkovnikov2011a,Firstenberg2013,Laird2015,Abanin2019}.
Dephasing has to be taken into account when designing mesoscopic devices
\citep{Clerk2010,Hammerer2010}, specifically those directed at quantum
information processing \citep{Makhlin2001a,Suter2016,Gambetta2006,FriskKockum2012}.

A particularly intriguing type of dephasing appears when geometrical
phases \citep{Berry1984,Cohen2019} emerge in open quantum systems
\citep{Ellinas1989,Gamliel1989,Gaitan1998a,Avron1998,Carollo2003,DeChiara2003a,Whitney2003,Whitney2005}.
On top of conventional dynamical dephasing (arising due to the fluctuations
of the system’s energy and proportional to the evolution time), Refs.~\citep{Whitney2003,Whitney2005,Whitney2006}
found a geometrical contribution to dephasing. Such geometrical dephasing
(GD) has two salient features. First, it can be expressed through
an integral of the underlying Berry curvature \citep{Snizhko2019b,Snizhko2019e}.
Second, similarly to Hamiltonian-generated geometrical phase, GD flips
its sign upon the reversal of the evolution protocol (the directionality
in which the closed path is traversed). The existence of geometrical
dephasing has been confirmed experimentally \citep{Berger2015}. Recent
theoretical studies \citep{Snizhko2019b,Snizhko2019e} have generalized
GD to the case of non-Abelian phases.

On a seemingly unrelated front, measurement-induced geometrical phases
have recently become an object of both experimental \citep{Cho2019}
and theoretical \citep{Gebhart2020} interest. Notably, measurement
in quantum mechanics involves stochasticity. It is thus natural to
ask whether dephasing emerges in measurement-based protocols \citep{DephasingFoot}\nocite{Gambetta2006,FriskKockum2012}
and to investigate its relation to Hamiltonian-induced dephasing \citep{AsymmetryFoot}.

\begin{figure}
\begin{centering}
\includegraphics[width=1\columnwidth]{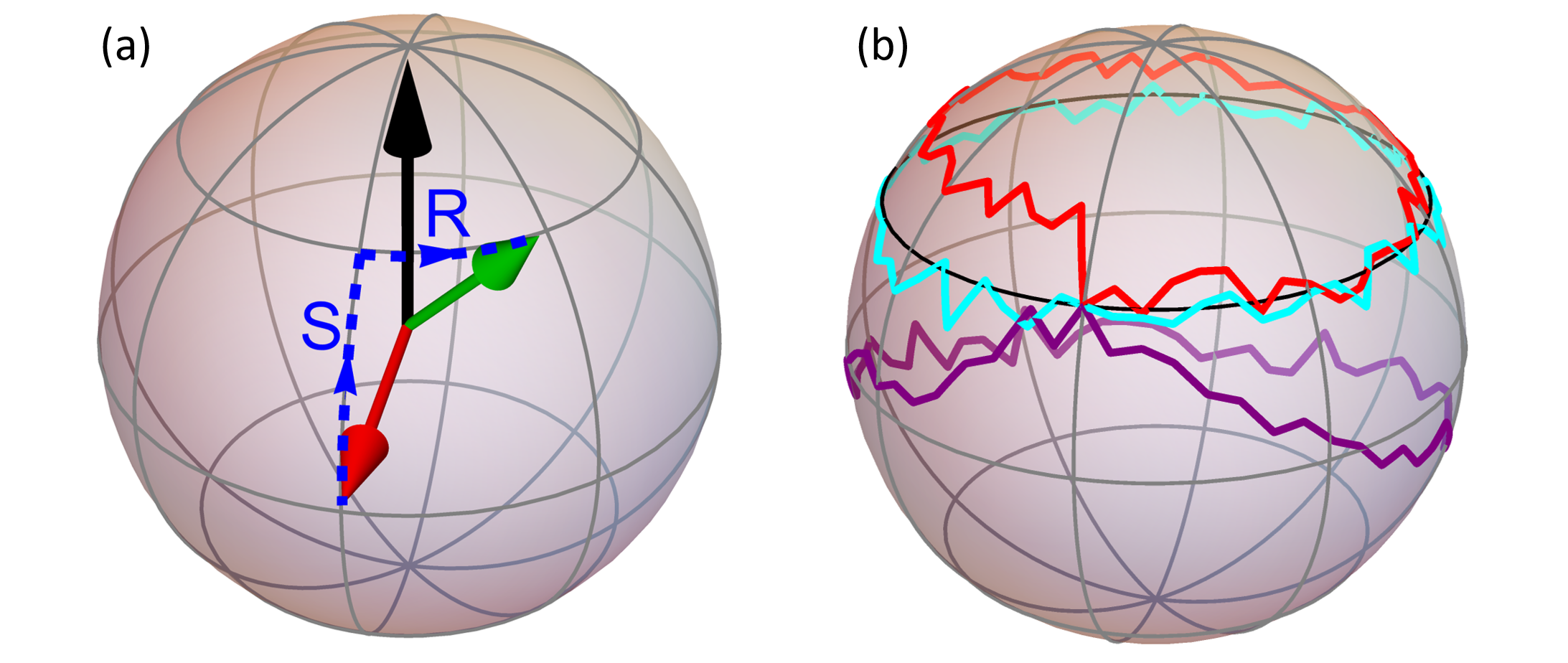}
\par\end{centering}
\caption{\label{fig:trajectories_on_the_Bloch_sphere}Measurement-induced trajectories.
(a)---The back-action of a generalized measurement. Under a projective
measurement yielding readout $r=0$, the system's initial state (red
arrow) would become $\protect\ket{\uparrow}$ (black arrow). For finite
measurement strength, the state is only pulled towards $\protect\ket{\uparrow}$
and also rotated around the $z$ axis (green arrow). These two back-action
aspects (illustrated by dashed blue lines) are quantified by parameters
$S$ and $R$ in Eq.~(\ref{eq:Kraus_C,A}). The direction of the
rotation (the sign of $R$) sets the measurement's chirality. (b)---Schematic
illustration of trajectories induced by a sequence of measurements
along some parallel (black line). Depending on the sequence of readouts
$\{r_{k}\}$, different trajectories (red, cyan, purple) and different
phases $\chi_{\{r_{k}\}}^{(d)}$ are induced.}
\end{figure}

The challenge of the present paper is two-fold. We first ask whether
weak-measurement-induced phases go hand-in-hand with emergent dephasing.
Secondly, provided that dephasing is part of such protocols, does
this dephasing have a term similar to GD? Our main findings are: (i)
Indeed, measurement-induced generation of phases does give rise to
dephasing. (ii) In similitude to Hamitonian dynamics of dissipative
systems, leading to dynamical and geometrical components, here both
the phase and the dephasing generated by measurement protocols comprise
a symmetric and an antisymmetric (w.r.t. changing directionality)
components. The latter manifests the existence of intrinsic chirality
in weak measurements (cf.~Fig.~\ref{fig:trajectories_on_the_Bloch_sphere}).
(iii) The emergent dephasing in such measurement-based steering protocols
may diverge. These divergences are associated with topological transitions
underlying the steering protocols.

\emph{Measurement model}.---As a concrete (but generalizable) example
we consider a spin-$1/2$ system subject to non-projective measurements.
Our detector is a two-level quantum object, its states labeled as
$r=0$, $1$. The measurement procedure comprises coupling the detector
to the system, decoupling it, and then measuring the detector projectively
in the $\ket r$ basis. The system evolution under measurement can
be described as $\ket{\psi}\rightarrow\mathcal{M}^{(r)}\ket{\psi}$
\citep{Nielsen2010,Wiseman2010,Jacobs2014a} where $\mathcal{M}^{(r)}$
is the generalized measurement operator (also known as the Kraus operator)
associated with the readout $r=0/1.$ 

Resorting to a protocol where we are blind to the detector readouts
(i.e., tracing out the detector), the measurement is indistinguishable
from a (carefully engineered) Markovian environment. However, as we
shall see in the coming sections, a non-conventional ``two-replica''
averaging over measurement readouts is important for the effects reported
here. 

Depending on the system-detector coupling, the measurement back-action
may vary. We focus on the specific case with $\mathcal{M}_{\mathbf{n}}^{(r)}=U^{-1}(\mathbf{n})M^{(r)}U(\mathbf{n})$,
where
\begin{equation}
U(\mathbf{n})=\begin{pmatrix}\cos\frac{\theta}{2} & \sin\frac{\theta}{2}e^{-i\varphi}\\
\sin\frac{\theta}{2} & -\cos\frac{\theta}{2}e^{-i\varphi}
\end{pmatrix},\label{eq:rotation_matrix}
\end{equation}
\begin{equation}
M^{(0)}=\begin{pmatrix}1 & 0\\
0 & e^{-S-iR}
\end{pmatrix},\quad M^{(1)}=\begin{pmatrix}0 & 0\\
0 & \sqrt{1-e^{-2S}}
\end{pmatrix},\label{eq:Kraus_C,A}
\end{equation}
with $(\theta,\varphi)$ being the spherical angles specifying the
measured spin direction $\mathbf{n}$.

The effect of such measurements is illustrated in Fig.~\ref{fig:trajectories_on_the_Bloch_sphere}(a).
Upon reading out $r=0$, the spin is pulled toward direction $\mathbf{n}$
(at strength $S$), and is rotated around it (cf.~the parameter $R$)
\citep{LambShiftFoot}. In conventional measurement models, this rotation
is disregarded. However, it can exist in practice, and, as is shown
below, it plays a crucial role in the behavior of measurement-induced
phases and dephasing. In particular, the sign of $R$ determines the
measurement's chirality. The readout $r=1$ results in the spin being
\emph{projected} onto direction $-\mathbf{n}$. The measurement strength
$S$ determines the probability $p_{r}=\bra{\psi}\mathcal{M}_{\mathbf{n}}^{(r)\dagger}\mathcal{M}_{\mathbf{n}}^{(r)}\ket{\psi}$
of this event. For a vanishing measurement strength, $S=0$, the readout
$r=1$ never occurs. The standard projective measurement is recovered
for $S\rightarrow\infty$. Note that in the limit of a projective
measurement $R$ has no effect, which is why the effect of measurement
chirality has been overlooked so far.

\emph{The protocol}.---Details of our protocol are inspired by studies
of dynamical quantum Zeno effect, cf.~Ref.~\citep{Facchi1999a},
where a spin-$1/2$ system is prepared in the direction $\mathbf{n}_{0}=(\sin\theta,0,\cos\theta)$
and then subjected to a sequence of $N+1$ projective measurements
corresponding to directions $\mathbf{n}_{k}=(\sin\theta\cos\varphi_{k},\sin\theta\sin\varphi_{k},\cos\theta)$,
$\varphi_{k}=2\pi kd/(N+1)$. Here $d=\pm1$ defines the directionality
of the trajectory. When $N\rightarrow\infty$, the spin state follows
the measurement directions with probability 1 and acquires the geometric
Pancharatnam phase $-\pi d(1-\cos\theta)$.

We employ the same protocol for inducing phases by measurements, but
with the first $N$ measurements being non-projective, as defined
in Eqs.~(\ref{eq:rotation_matrix}--\ref{eq:Kraus_C,A}) with $S=2C/N$
and $R=2A/N$. The parameters $C$ and $A$ (of $O(1)$) characterize
the measurement; the $1/N$ scaling is required in order to avoid
the quantum Zeno effect in the $N\rightarrow\infty$ limit, allowing
for non-trivial evolution of the system's trajectory and the study
of measurement chirality. With this modification, the spin state follows
a trajectory parameterized by the readout sequence $\{r_{k}\}$, cf.~Fig.~\ref{fig:trajectories_on_the_Bloch_sphere}(b)
(each sequence is associated with a specific probability). We keep
the final measurement, $k=N+1$, projective, and postselect it on
yielding $r=0$ readout in order to ensure that the spin has followed
a closed trajectory on the Bloch sphere.

\emph{The observable}.---For each readout sequence $\{r_{k}\}$,
the spin state acquires a phase $\chi_{\{r_{k}\}}^{(d)}$. After averaging
over different readout sequences, we define the averaged phase, $\bar{\chi}^{(d)}$,
and the dephasing parameter, $\alpha^{(d)}$, through
\begin{equation}
\langle e^{2i\chi_{\{r_{k}\}}^{(d)}}\rangle_{\{r_{k}\}}=e^{2i\bar{\chi}^{(d)}-\alpha^{(d)}}.\label{eq:av_phase_def}
\end{equation}

This somewhat arbitrary form of averaging in Eq.~(\ref{eq:av_phase_def})
is motivated by the following proposal for observing the measurement-induced
phases \citep{Gebhart2020,Snizhko2020c}. The phase could, naively,
be measured by an interference experiment, where a ``flying spin-1/2'',
represented by an impinging electron, is split between two arms and
subjected to measurements in one of them. Such a protocol, however,
presents the following problem: the detector changing its state would
not only induce a back-action on the system (the flying spin-1/2),
but would also constitute a ``which-path'' measurement, undermining
the interference, just as conventional coupling to the environment
would. Instead, one can resort to a measurement setup where each detector
is coupled to respective points on both arms. The system-detector
couplings are engineered such that the phases accumulated in the respective
arms are $\chi_{\{r_{k}\}}^{(d)}$ and $-\chi_{\{r_{k}\}}^{(d)}$
for each and every sequence of readouts $\{r_{k}\}$. With these designed
couplings, the probabilities of obtaining a specific readout sequence
$\{r_{k}\}$ are identical for both arms, hence, no ``which path''
measurement. The interference pattern then corresponds to the relative
phase $e^{2i\chi_{\{r_{k}\}}^{(d)}}$ and is averaged over runs with
different readout sequences $\{r_{k}\}$.

\emph{Derivation of the dephasing factor}.---The phase accumulated
under a sequence of measurements can be calculated as follows. Denote
the initial system state $\ket{\psi_{0}}=\cos\frac{\theta}{2}\ket{\uparrow}+\sin\frac{\theta}{2}\ket{\downarrow}$.
After performing the sequence of $N$ generalized measurements, for
a given readout sequence $\{r_{k}\}=\{r_{1},...,r_{N}\}$, the system
state becomes $\mathcal{M}_{\mathbf{n}_{N}}^{(r_{N})}...\mathcal{M}_{\mathbf{n}_{2}}^{(r_{2})}\mathcal{M}_{\mathbf{n}_{1}}^{(r_{1})}\ket{\psi_{0}}$.
The last projective measurement makes the system state $\ket{\psi_{0}}\bra{\psi_{0}}\mathcal{M}_{\mathbf{n}_{N}}^{(r_{N})}...\mathcal{M}_{\mathbf{n}_{1}}^{(r_{1})}\ket{\psi_{0}}$.
The matrix element 
\begin{equation}
\bra{\psi_{0}}\mathcal{M}_{\mathbf{n}_{N}}^{(r_{N})}...\mathcal{M}_{\mathbf{n}_{1}}^{(r_{1})}\ket{\psi_{0}}=\sqrt{P_{\{r_{k}\}}^{(d)}}e^{i\chi_{\{r_{k}\}}^{(d)}}\label{eq:chi_definition}
\end{equation}
defines the measurement-induced phase $\chi_{\{r_{k}\}}^{(d)}$ and
the probability $P_{\{r_{k}\}}^{(d)}$ of obtaining readout sequence
$\{r_{k}\}$ (including $r=0$ for the last \emph{projective} measurement,
bringing the system to $\ket{\psi_{0}}$). Considering all possible
measurement readout sequences $\{r_{k}\}$, the averaged phase $\bar{\chi}^{(d)}$
and the dephasing parameter $\alpha^{(d)}$ are given by 
\begin{multline}
e^{2i\bar{\chi}^{(d)}-\alpha^{(d)}}=\sum_{\{r_{k}\}}\left(\bra{\psi_{0}}\mathcal{M}_{\mathbf{n}_{N}}^{(r_{N})}...\mathcal{M}_{\mathbf{n}_{1}}^{(r_{1})}\ket{\psi_{0}}\right)^{2}\\
=\sum_{\{r_{k}\}}P_{\{r_{k}\}}^{(d)}e^{2i\chi_{\{r_{k}\}}^{(d)}}.\label{eq:chi-bar_definition}
\end{multline}

We compute $e^{2i\bar{\chi}^{(d)}-\alpha^{(d)}}$ using the following
trick. Note that $\bra{\psi_{0}}\mathcal{M}_{\mathbf{n}_{N}}^{(r_{N})}...\mathcal{M}_{\mathbf{n}_{1}}^{(r_{1})}\ket{\psi_{0}}=\bra{\uparrow}\delta UM^{(r_{N})}\delta U...\delta UM^{(r_{1})}\delta U\ket{\uparrow}$,
where $\delta U=U(\mathbf{n}_{k+1})U^{-1}(\mathbf{n}_{k})$ is a matrix
that does not depend on $k$, cf.~Eqs.~(\ref{eq:rotation_matrix}--\ref{eq:Kraus_C,A}).
In order to calculate the sum over $\{r_{k}\}$, we define a matrix
$\mathfrak{M}_{s_{1}s_{2}}^{s_{1}'s_{2}'}=\sum_{r}\bra{s_{1}'}M^{(r)}\delta R\ket{s_{1}}\bra{s_{2}'}M^{(r)}\delta R\ket{s_{2}}$.
Here $s_{i}$ (``before the measurement'') and $s_{i}'$ (``after
the measurement'') take values $\uparrow/\downarrow$ with $i=1,2$
being the replica index. We find that for $N\rightarrow\infty$
\begin{equation}
\mathfrak{M}=\mathbb{I}+\Lambda/N+O(N^{-2}),\label{eq:M_largeN}
\end{equation}
where $\Lambda$ is a constant matrix that depends on the measurement
parameters $C$ and $A$, polar angle $\theta$, and directionality
$d$, characterizing the protocol:\begin{widetext}
\begin{equation}
\Lambda=\begin{pmatrix}2i\pi d\cos\theta & -i\pi d\sin\theta & -i\pi d\sin\theta & 0\\
-i\pi d\sin\theta & -2(C+iA) & 0 & -i\pi d\sin\theta\\
-i\pi d\sin\theta & 0 & -2(C+iA) & -i\pi d\sin\theta\\
0 & -i\pi d\sin\theta & -i\pi d\sin\theta & -2i\pi d\cos\theta-4iA
\end{pmatrix}\begin{bmatrix}\uparrow\uparrow\\
\uparrow\downarrow\\
\downarrow\uparrow\\
\downarrow\downarrow
\end{bmatrix}.\label{eq:Lambda4x4}
\end{equation}
\end{widetext} Then
\begin{equation}
e^{2i\bar{\chi}^{(d)}-\alpha^{(d)}}=\lim_{N\rightarrow\infty}\left(\mathfrak{M}^{N}\right)_{\uparrow\uparrow}^{\uparrow\uparrow}=\left[\exp(\Lambda)\right]_{\uparrow\uparrow}^{\uparrow\uparrow}.\label{eq:av_phase_via_M4x4^N}
\end{equation}

\emph{Classification of dephasing.}---In light of Eq.~(\ref{eq:av_phase_via_M4x4^N}),
it is tempting to denote $\alpha^{(d)}$ and $\bar{\chi}^{(d)}$ geometrical
since they do not depend on the protocol duration (number of measurements).
Such an identification would be erroneous, as can be easily seen from
the following argument: For $C=0$ (equivalently, $S=0$ in Eq.~(\ref{eq:Kraus_C,A})),
our measurement-induced evolution is equivalent to \emph{non-adiabatic}
Hamiltonian evolution (readout $r=1$ never occurs; the back-action
with the Kraus operator $\mathcal{M}_{\mathbf{n}}^{(0)}$ is equivalent
to a Hamiltonian rotation). While the accumulated phase in that case
does not scale with $N$, it is known that it admits a non-trivial
separation into the dynamical and geometrical components \cite{Aharonov1987, [See also ] Wood2020}.
At the same time, these dynamical and geometrical components behave
non-trivially with respect to directionality reversal, $d\rightarrow-d$,
which hinders a simple classification based on symmetry properties.
Here we do not delve deeper into this classification issue \citep{LongPaperFoot}
but rather focus on the behavior of dephasing and the measurement-induced
phase in the context of directionality reversal.

To understand the relation between $\alpha^{(d)}$ and $\alpha^{(-d)}$,
we note the following symmetries of $\Lambda$. From Eq.~(\ref{eq:Lambda4x4}),
we see that replacing $d\rightarrow-d$, $A\rightarrow-A$, together
with a complex conjugation, leaves $\Lambda$ invariant, i.e. $\Lambda_{d\rightarrow-d,A\rightarrow-A}=\Lambda^{*}$.
Using Eq.~(\ref{eq:av_phase_via_M4x4^N}), this implies $\alpha^{(d)}(C,A,\theta)=\alpha^{(-d)}(C,-A,\theta)$
and $\bar{\chi}^{(d)}(C,A,\theta)=-\bar{\chi}^{(-d)}(C,-A,\theta)$.
Consequently, the dephasing is only guaranteed to be symmetric ($\alpha^{(d)}=\alpha^{(-d)}$)
when $A=0$. Away from $A=0$ there may be an additional antisymmetric
component. We therefore denote $A$ as the asymmetry parameter. Using
the above symmetry relations, we write down the symmetric and antisymmetric
dephasing components, $\alpha^{(d)}=\alpha^{s}+\alpha^{a}d$,
\begin{align}
\alpha^{s} & =\frac{1}{2}\left(\alpha^{(+1)}(C,A,\theta)+\alpha^{(+1)}(C,-A,\theta)\right),\label{eq:sym_deph_through_+1}\\
\alpha^{a} & =\frac{1}{2}\left(\alpha^{(+1)}(C,A,\theta)-\alpha^{(+1)}(C,-A,\theta)\right).\label{eq:antisym_deph_through_+1}
\end{align}

Next, defining a diagonal matrix $U=\mathrm{diag}(1,-1,-1,1)$, one
shows that $\Lambda_{d\rightarrow-d,\theta\rightarrow\pi-\theta}=U\Lambda U$,
which implies $\alpha^{(d)}(C,A,\theta)=\alpha^{(-d)}(C,A,\pi-\theta)$
and $\bar{\chi}^{(d)}(C,A,\theta)=\bar{\chi}^{(-d)}(C,A,\pi-\theta)$.
This symmetry can be understood from a simple consideration: a clockwise
($d=-1$) protocol in the southern hemisphere becomes a counterclockwise
($d=1$) protocol in the northern hemisphere upon exchanging the roles
of the south and the north poles.

\begin{figure}
\begin{centering}
\includegraphics[width=1\columnwidth]{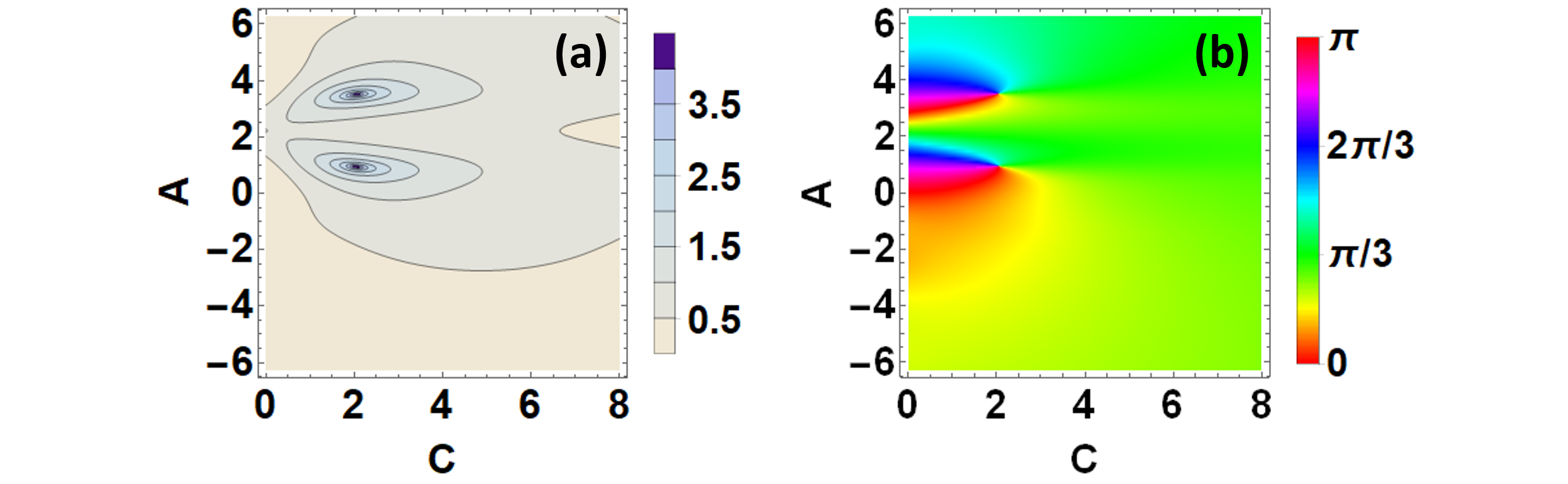}
\par\end{centering}
\caption{\label{fig:avg_dephasing_and_phase}Dephasing $\alpha^{(+1)}$ (a)
and phase $\bar{\chi}^{(+1)}$ (b), cf.~Eq.~(\ref{eq:av_phase_def}),
at $\theta=3\pi/4$ color-coded as functions of the measurement strength
($C$) and asymmetry ($A$) parameters. Note the difference in behavior
for $A>0$ and $A<0$, showing the effect of the measurement chirality.
Most prominent are the two singularities at $C\approx2$, where $\alpha^{(+1)}$
diverges. The phase makes $\pi$-windings around the points of divergent
$\alpha^{(+1)}$.}
\end{figure}

\emph{Asymmetric dephasing}.---We next calculate numerically and
analyze the behavior of $\alpha^{(+1)}$ (the behavior of $\alpha^{(-1)}$
can be inferred by swapping $\theta\rightarrow\pi-\theta$). Figure~\ref{fig:avg_dephasing_and_phase}(a)
shows the dependence of $\alpha^{(+1)}$ on the measurement parameters
$C$ and $A$ at $\theta=3\pi/4$. Note that $\alpha^{(+1)}(C,A,\theta)\neq\alpha^{(+1)}(C,-A,\theta)=\alpha^{(-1)}(C,A,\theta)$,
revealing that the antisymmetric component $\alpha^{a}$ is indeed
generically present as soon as $A\neq0$ (i.e., whenever the measurements
employed in the protocol possess chirality).

Note also the two divergences, $\alpha^{(+1)}\rightarrow\infty$,
at $(C_{\mathrm{crit}}\approx2,A_{\mathrm{crit}}>0)$. There are no
corresponding divergences at $A<0$, implying that $\alpha^{(-1)}(C_{\mathrm{crit}},A_{\mathrm{crit}},\theta)$
is non-singular. This implies that both $\alpha^{a}(C_{\mathrm{crit}},A_{\mathrm{crit}},\theta)$
and $\alpha^{s}(C_{\mathrm{crit}},A_{\mathrm{crit}},\theta)$ diverge.
Moreover, the strength of divergence is identical as $\alpha^{(-1)}=\alpha^{s}-\alpha^{a}$
is finite. Contrast this to the case of dephasing in Hamiltonian-induced
dynamics, where $\alpha^{(d)}=\beta\mathcal{E}T+\gamma d$ with $\mathcal{E}T\gg1$
being the adiabaticity parameter ($\mathcal{E}$ is the energy gap
and $T$ is the protocol execution time) \citep{Whitney2003,Whitney2005,Berger2015}.
There the symmetric component, $\alpha^{s}=\beta\mathcal{E}T$, associated
with dynamical dephasing, always dominates over the antisymmetric
geometrical dephasing $\alpha^{a}=\gamma d$ so that $\alpha^{s}/\alpha^{a}\sim\mathcal{E}T\gg1$.

\emph{Divergences as topological features}.---Consider the phase,
$\bar{\chi}^{(+1)}(\theta=3\pi/4)$, whose dependence on $C$ and
$A$ is shown in Fig.~\ref{fig:avg_dephasing_and_phase}(b). Remarkably,
the phase makes a winding around each of the dephasing singularity
points, cf.~Fig.~\ref{fig:avg_dephasing_and_phase}(a). Exactly
at the singularity points, the phase is undefined since $e^{2i\bar{\chi}^{(+1)}-\alpha^{(+1)}}=0$.
The windings of the phase are of size $\pi$, and not $2\pi$. However,
since the measurable quantity is $e^{2i\bar{\chi}^{(+1)}}$, there
is no physical discontinuity as $\bar{\chi}^{(+1)}\rightarrow\bar{\chi}^{(+1)}+\pi$.
The windings cannot be eliminated by a continuous deformation of the
phase, and thus constitute topological features. Such phase windings
accompany all the divergences we found.

\begin{figure}
\begin{centering}
\includegraphics[width=1\columnwidth]{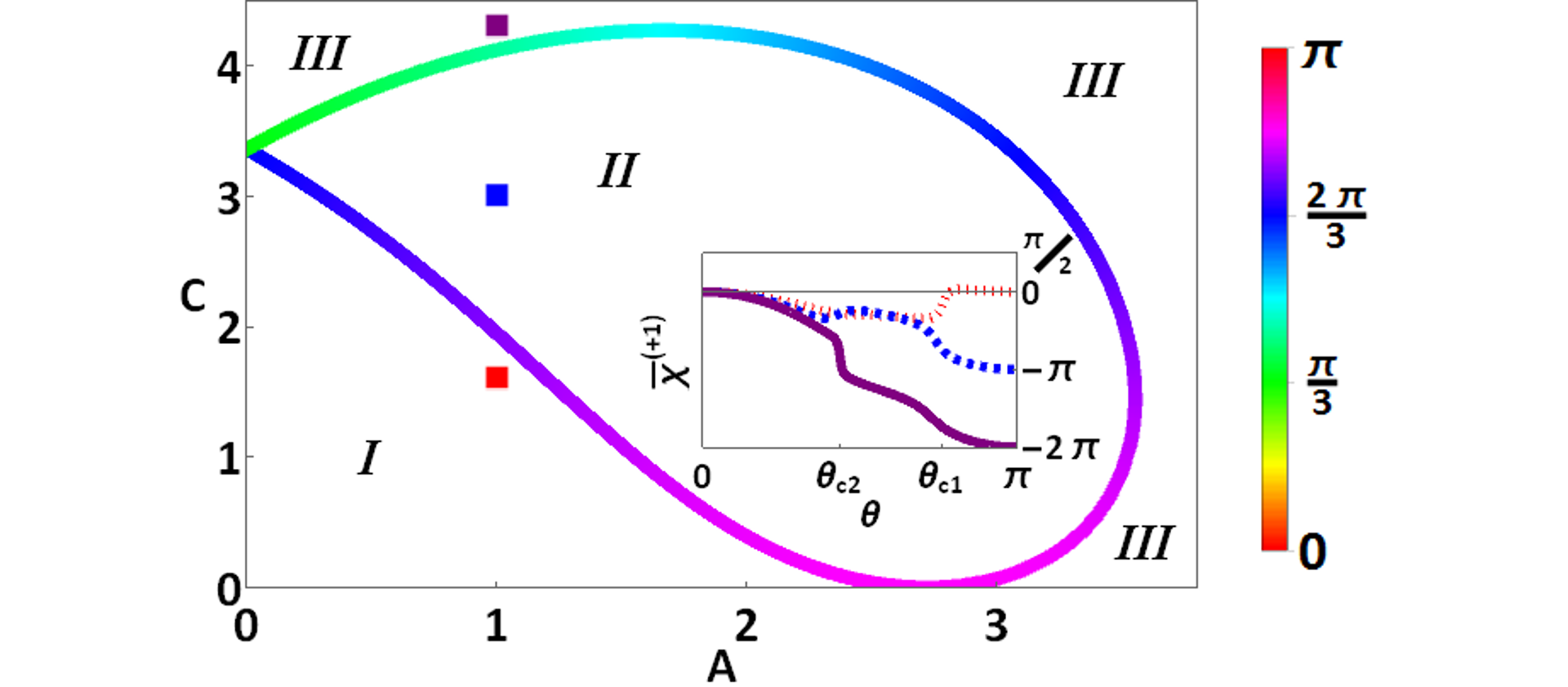}
\par\end{centering}
\caption{\label{fig:phase_diagram}Topological transition in the measurement-induced
phase $\bar{\chi}^{(+1)}$. Main panel---The critical line of points
$(C_{\mathrm{crit}},A_{\mathrm{crit}})$ for which there exists $\theta_{\mathrm{crit}}$
such that the dephasing $\alpha^{(+1)}(C_{\mathrm{crit}},A_{\mathrm{crit}},\theta_{\mathrm{crit}})$
diverges. The values of $\theta_{\mathrm{crit}}\in[0;\pi]$ are shown
with color. The averaged phase $\bar{\chi}^{(+1)}$ exhibits three
distinctly different topological behaviors, corresponding to regions
$I$, $II$, and $III$. Inset---Dependence of $\bar{\chi}^{(+1)}$
on $\theta$ for the measurement parameters $(C,A)$ marked with squares
in the main plot. As $\theta$ varies from $0$ to $\pi$, $\bar{\chi}^{(+1)}(\theta)$
varies from $0$ to $0$ (region $I$), $0$ to $-\pi$ (reigon $II$)
, or $0$ to $-2\pi$ (region $III$). The values of $\theta_{\mathrm{crit}}$
corresponding to the two transitions at $A=1$ are marked as $\theta_{\mathrm{c}1}$
and $\theta_{\mathrm{c}2}$.}
\end{figure}

Another way of viewing the divergences as topological features arises
when considering the set of all divergences. The divergences of $\alpha^{(+1)}$
form a critical line $(C_{\mathrm{crit}},A_{\mathrm{crit}})$ in the
$(C,A)$ plane, cf.~Figure~\ref{fig:phase_diagram}. For each $(C_{\mathrm{crit}},A_{\mathrm{crit}})$,
there is a value of $\theta_{\mathrm{crit}}\in[0;\pi]$ at which $\alpha^{(+1)}$
diverges. The critical line separates the plane into three regions.
The $\theta$-dependence of the phase $\bar{\chi}^{(+1)}(C,A,\theta)$
is topologically different in each of these regions. To see this,
consider the dependence on the polar angle, $\theta$, of $\bar{\chi}^{(+1)}(\theta)$
for a given value of measurement parameters $(C,A)$. For each given
$\theta$, $\bar{\chi}^{(+1)}$ is defined modulo $\pi$. However,
taking the whole dependence on $\theta$ into account, we unfold the
phase to form a continuous function $\bar{\chi}^{(+1)}(\theta)$ which
is not confined to the interval $[0;\pi)$. Furthermore, note that
$e^{2i\bar{\chi}^{(+1)}(\theta=0)}=e^{2i\bar{\chi}^{(+1)}(\theta=\pi)}=1$.
This implies
\begin{equation}
\bar{\chi}^{(+1)}(\pi)=\bar{\chi}^{(+1)}(\pi)-\bar{\chi}^{(+1)}(0)=\int_{0}^{\pi}d\theta\frac{d\bar{\chi}^{(+1)}(\theta)}{d\theta}=\pi\bar{n},
\end{equation}
where $\bar{n}\in\mathbb{Z}$ and we have used the freedom to fix
$\bar{\chi}^{(+1)}(0)=0$. No transition between different values
of integer $\bar{n}$ can happen when $\bar{\chi}^{(+1)}(\theta)$
is smoothly deformed, making $\bar{n}$ a topological index. However,
$\bar{n}(C,A)$ can jump when $\bar{\chi}^{(+1)}(\theta)$ is not
a well-defined smooth function. This happens at the divergence points
where $\alpha^{(d)}(C_{\mathrm{crit}},A_{\mathrm{crit}},\theta_{\mathrm{crit}})=+\infty$,
so that $\bar{\chi}^{(+1)}(C_{\mathrm{crit}},A_{\mathrm{crit}},\theta_{\mathrm{crit}})$
is undefined. Therefore, the $(C,A)$ plane can be divided into regions,
each with a distinct value of $\bar{n}$. In the present example,
$\bar{n}=0$ in region $I$, $\bar{n}=-1$ inside $II$, and $\bar{n}=-2$
inside $III$, as illustrated in Fig.~\ref{fig:phase_diagram}(inset).
The behavior of $\bar{\chi}^{(-1)}(\theta)$ is recovered via relation
$\bar{\chi}^{(d)}(C,A,\theta)=\bar{\chi}^{(-d)}(C,A,\pi-\theta)(\mathrm{mod}\,\pi)$,
implying that for $d=-1$ similar topological transitions happen at
the same $(C_{\mathrm{crit}},A_{\mathrm{crit}})$ but at different
$\theta_{\mathrm{crit}}$.

We emphasize the peculiarity of region $II$. Note that the phases
$\chi_{\{r_{k}\}}^{(d)}$ corresponding to individual readout sequences
are defined modulo $2\pi$, cf.~Eq.~(\ref{eq:chi_definition}).
Therefore, the behavior of regions $I$ and $III$ can be anticipated
and observed for individual readout sequences \citep{LongPaperFoot}.
However, no individual readout sequence allows for $\chi_{\{r_{k}\}}^{(d)}(\pi)-\chi_{\{r_{k}\}}^{(d)}(0)=\pm\pi$.
Thus, the behavior of region $II$ is a highly non-trivial consequence
of averaging over multiple readout sequences in Eq.~(\ref{eq:av_phase_def}).
This is reminiscent of the relation between the fields of dissipative
topological matter \citep{Kraus2008,Weimer2010,Diehl2011,Henriet2017,Roy2020a}
and measurement-induced entanglement transitions \citep{Li2018a,Chan2019,Skinner2019a,Li2019a,Gullans2019,Cao2019a,Szyniszewski2019,Szyniszewski2020,Tang2020c,Choi2020c,Ippoliti2020,Li2020o,Buchhold2021}.
In the latter, looking at a refined observable that essentially depends
on a non-trivial readout-averaging, gives rise to a number of new
effects \citep{Bao2021a}. Furthermore, region $II$ is only present
when $A\neq0$, showing that this behavior crucially depends on the
measurement exhibiting chirality.

\emph{Summary}.---We have presented here a protocol comprising a
set of generalized measurements, which steers a spin-1/2 system along
a closed trajectory on the Bloch sphere. Fluctuations in the readout
sequences are responsible for dephasing, which is not symmetric under
changing of path directionality, $d\rightarrow-d$. Rather it comprises
two components: symmetric and antisymmetric. The latter is a manifestation
of the measurement's intrinsic chirality, a feature which, to the
best of our knowledge, has not been previously emphasized.

The measurement-induced dynamics bears similitude to adiabatic Hamiltonian
dynamics of open quantum systems, where symmetric (dynamical) and
antisymmetric (geometrical) dephasing components have been predicted
and observed \citep{Whitney2003,Whitney2005,Whitney2006,Berger2015}.
Indeed, the detector can be thought of as an external environment,
while initializing the detector before each measurement amounts to
Markovianity, often implied when describing open systems. At the same
time, we find a number of important differences between these two
paradigms of dephasing. For adiabatic Hamiltonian dynamics the identification
of the symmetric and antisymmetric components with dynamical and geometrical
dephasing respectively is clear-cut. This is not the case with measurement-induced
dynamics. Furthermore, while in adiabatic Hamiltonian dynamics the
antisymmetric component is always \emph{much} smaller than the symmetric
one, this does not apply for measurement-induced dephasing (nevertheless,
the symmetric term always exceeds the antisymmetric term, which guarantees
that for either directionality $d$ the overall effect is suppression
of coherent terms).

We have found divergences of the measurement-induced dephasing and
linked them to topological transitions in the behavior of the measurement-induced
phase factors. We note that a special case of such a transition ($A=0$,
cf.~Eq.~\ref{eq:Kraus_C,A}) has been discovered in Ref.~\citep{Gebhart2020}.
We thus conclude that such transitions are a richer phenomenon than
previously thought. This is revealed by the corresponding ``phase
diagram'', cf.~Fig.~\ref{fig:phase_diagram}. In particular, region
$II$ in this phase diagram is only present for $A\neq0$, i.e., for
measurements that are chiral.

Finally, we stress that our findings extend beyond the concrete protocol
and the specific type of measurements studied here. In particular,
measurement-induced phases exhibit dephasing for an arbitrary number
of measurements ($N<\infty$), arbitrary Kraus operators ($\mathcal{M}^{(r)}$),
and arbitrary sequences of measurement directions ($\mathbf{n}_{k}$).
The dephasing will, in general, be asymmetric w.r.t. reversal of the
protocol directionality, and may diverge under certain conditions.
For $N<\infty$, the dephasing and the induced phase will depend on
$N$.

\begin{acknowledgments}

We thank V. Gebhart for useful discussions. We acknowledge funding
by the Deutsche Forschungsgemeinschaft (DFG, German Research Foundation)
-- Projektnummer 277101999 -- TRR 183 (project C01) and Projektnummern
EG 96/13-1, GO 1405/6-1, and MI 658/10-2, and by the Israel Science
Foundation (ISF).

\end{acknowledgments}

\bibliography{bibliography,extra}

\end{document}